\documentclass{article}
\usepackage{arxiv}
\usepackage[utf8]{inputenc} 
\usepackage[T1]{fontenc}    
\usepackage[colorlinks=true, allcolors=blue]{hyperref}       
\usepackage{url}            
\usepackage{booktabs}       
\usepackage{amsfonts}       
\usepackage{nicefrac}       
\usepackage{microtype}      
\usepackage{lipsum}		
\usepackage{graphicx}
\usepackage{natbib}
\usepackage{doi}
\usepackage{adjustbox}
\usepackage[version=4]{mhchem}

\title{ALMA detection of water vapour in the low mass protostar IRAS 16293--2422}

\author{ {\color{red}Arijit Manna}\\
	Midnapore City College\\
	Kuturia, Bhadutala, Paschim Medinipur, \\West Bengal, India 721129 \\
	\texttt{mannaarijit@hotmail.com} \\
	\And
	{\color{red}Sabyasachi Pal} \\
	Indian Centre for Space Physics\\43 Chalantika, Garia Station Road, \\Kolkata, India 700084\\\\
	Midnapore City College\\
	Kuturia, Bhadutala, Paschim Medinipur, \\West Bengal, India 721129 \\
	\texttt{sabya.pal@gmail.com} \\
	\And
	{\color{red}Soumyadip Banerjee} \\
	Midnapore City College\\
	Kuturia, Bhadutala, Paschim Medinipur, \\West Bengal, India 721129 \\
	\texttt{soumyadipbanerjee@hotmail.com}\\
}

\renewcommand{\shorttitle}{\ce{H2O} in IRAS 16293-2422}

\begin{document}
\maketitle

\begin{abstract}
	The low mass protostar IRAS 16293--2422 is well-known young stellar system that is observed in the L1689N molecular cloud in the constellation of Ophiuchus. In the interstellar medium and solar system bodies, water is a necessary species for the formation of life. We present the spectroscopic detection of the rotational emission line of water (\ce{H2O}) vapour from the low mass protostar IRAS 16293--2422 using the Atacama Large Millimeter/submillimeter Array (ALMA) band 5 observation. The emission line of \ce{H2O} is detected at frequency $\nu$ = 183.310 GHz with transition J=3$_{1,3}$--2$_{2,2}$. The statistical column density of the emission line of water vapour is $N$(\ce{H2O}) = 4.2$\times$10$^{16}$ cm$^{-2}$ with excitation temperature ($T_{ex}$) = 124$\pm$10 K. The fractional abundance of \ce{H2O} with respect to \ce{H2} is 1.44$\times$10$^{-7}$  where $N$(\ce{H2}) = 2.9$\times$10$^{23}$ cm$^{-2}$.
\end{abstract}


	\section{Introduction}
\label{sec:intro} 
The hot molecular cores are mainly found around the protostars which are characterized by the high density ($n \geq$ 10$^{16}$ cm$^{-3}$) and warm temperature ($T \geq$ 100 K). These types of protostars are mainly found in the hot cores around the Orion nebula. The complex and saturated molecular species are found in these types of protostars and these saturated molecular species are not abundant in the dark molecular cloud \citep{wal89}. The warm dense gas has been recently observed around the solar-type protostar IRAS 16293--2442 \citep{cec02}, which was historically identified only in the massive star formation region \citep{ban21,rav19,van19}. The low mass protostar IRAS 16293--2442 has been located in the $\rho$ Ophiuchus at the distance of 120 pc \citep{kun98}. 
The low mass protostar IRAS 16293--2442 consists of two cores which are called IRAS 16293A and IRAS 16293B and it is separated by $\sim$5$^{\prime\prime}$ \citep{woo89}. Earlier, several molecular outflows were detected in this protostar \citep{cas01,sta04,cha05,yeh08}. The hot corino has been first discovered in this class 0 protostar \citep{caz03,bot04}.

The water molecule is essential in the process of star formation by cooling warm gas, in addition to being a primordial component in the emergence of life. It also regulates the chemistry of a variety of species in the gas phase or on grain surfaces. The maser water emission is observed many times towards the low mass protostar IRAS 16293--2422. Earlier, \citet{im07} reported the detection of maser \ce{H2O} emission towards the IRAS 16293--2422 using VLBI Exploration of Radio Astrometry (VERA) array. The HDO/\ce{H2O} ratio must be high if water was formed at low temperatures, such as on cold grain surfaces, and it was created in the photodissociation region or by shock chemistry. The difference of the energy between \ce{H2O} and HDO is 886 K \citep{he05}. As a result, deuterated water enrichment with respect to its primary isotopologue occurs at low temperatures. In high-mass hot cores, the deuteration fraction is normally found HDO/\ce{H2O} $\leq$ 10$^{-3}$ \citep{jac90,gen96,hel96}, however higher values in order of $\sim$10$^{-2}$ has been recently discovered in Orion nebula \citep{per07}. The ratio of HDO/\ce{H2O} also has been calculated in the inner envelope of class 0 protostar such as IRAS 16293--2422 with upper limit 3\% \citep{par05}, NGC 1333-IRAS2A with a lower limit of 1\% \citep{lu11}, and NGC 1333-IRAS4B with upper limit 0.06\% \citep{jp10}.

\begin{figure}
	\centering
	\includegraphics[width=0.8\textwidth]{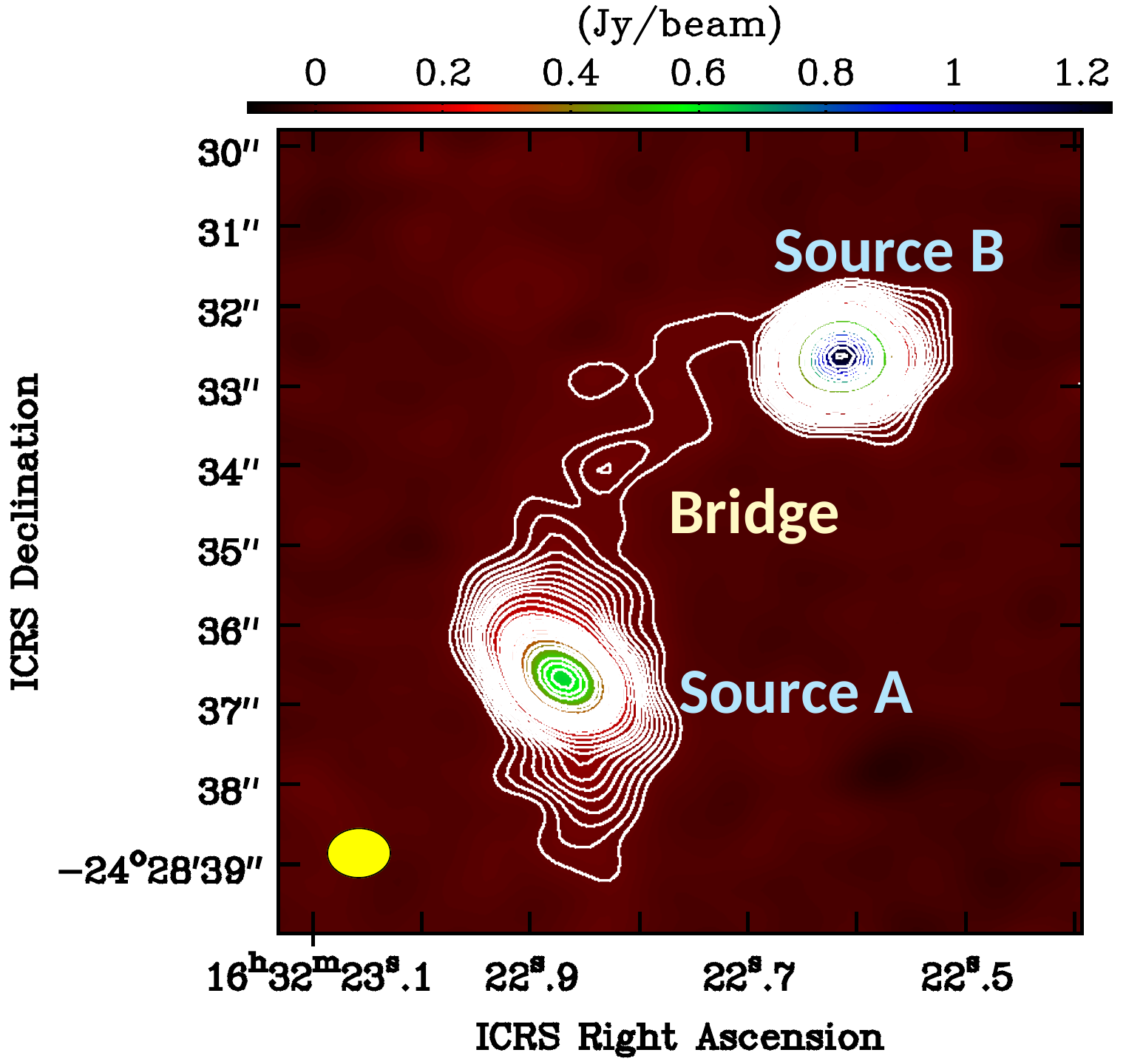}
	\caption{Radio continuum image of IRAS 16293--2422 with source A and B which is obtained at frequency 183.310 GHz with ALMA band 5. The synthesized beam is shown in the lower left corner with yellow colour which size is 0.39$^{\prime\prime}$$\times$0.32$^{\prime\prime}$. The relative contour levels set at 10 mJy beam$^{-1}$ (3$\sigma$) and increasing by the factor of $\surd$2. The peak continuum flux densities of source A and B is 0.25 and 0.59 Jy beam$^{-1}$ respectively.}
	\label{fig:cont}
\end{figure}

Many attempts to measure the fractionation of water deuterium in protostars have provided a variety of results. Earlier, the ground-based infrared observations of OD and OH in the outer parts of several low mass protostar were used to calculate the upper limits of HDO/\ce{H2O} ratio in the order of 0.5\% to 2\% \citep{par03}. \citet{cou12} also found the HDO/\ce{H2O} ratio in the order of 3.4$\times$10$^{-2}$ and 0.5$\times$10$^{-2}$ in the inner and outer envelops of IRAS 16293--2422 using the Herschel observation.

In this letter, we present the spectroscopic detection of water vapour at $\nu$ = 183.310 GHz toward IRAS 16293--2422 and we derive the excitation temperature of \ce{H2O} using LTE model to calculate the column density and to find relative abundance of \ce{H2O} with respect to \ce{H2} i.e, \ce{H2O}/\ce{H2} ratio. The observation and data reduction procedure is present in Sect.~\ref{obs}. The result and discussion of the detection of \ce{H2O} and modeling to find the column abundance as well as relative abundance is presented in Sect.~\ref{res}. The summary is presented in Sect.~\ref{sec:discussion}.

\section{Observations and data reduction}
\label{obs} 
The low mass protostar IRAS 16293--2422 was observed on July 10, 2018, using Atacama Large Millimeter/submillimeter Array (ALMA)\footnote{\href{https://almascience.nao.ac.jp/asax/}{https://almascience.nao.ac.jp/asax/}} with band 5. The phase center of the observation was ($\alpha,\delta$)$_{\rm J2000}$ = (16:32:22.720, --24:28:34.300). The spectral configuration was set up to observe the \ce{H2O} line at 183.310 GHz with a bandwidth of 0.059 GHz. On the observation date, the atmosphere condition was very good with precipitable water vapour (PWV) 0.2 mm. The observation was carried out using the forty-four number of antennas.

We used the Common Astronomy Software Application ({\tt CASA} 5.4.1)\footnote{\href{http://casa.nrao.edu/}{http://casa.nrao.edu/}} for initial data reduction and imaging of low mass protostar IRAS 16293--2422. We used the Perley-Butler 2017 \citep{per17} flux calibrator model for the absolute flux calibration. During the observation, J1625--2527 was observed as a flux calibrator, and J1633--2557 was observed as a phase calibrator. For initial data reduction, we apply the flux calibration and bandpass calibration using the flux calibrator. The details about the analysis of IRAS 16293--2422 are shown in {\tt CASA guide}\footnote{\href{https://casaguides.nrao.edu/index.php?title=ALMAguides}{https://casaguides.nrao.edu/index.php?title=ALMAguides}}. After the initial data reduction, we apply the task {\tt mstransform} to split the data set into calibrated target data set. The continuum image of IRAS 16293--2422 was created by the line-free channels between the frequency range 183.28--183.34 GHz using task {\tt tclean}. The self calibrated continuum image of IRAS 16293--2422 are shown in Fig.~\ref{fig:cont} in which synthesized beam size is 0.39$^{\prime\prime}$$\times$0.32$^{\prime\prime}$. After the creation of the continuum image, we apply the task {\tt uvcontsub} for the continuum subtraction. After the continuum subtraction, we create the emission map of IRAS 16293--2422 and extract the emission line of water vapour at the frequency 183.310 GHz.
\begin{figure}
			\centering
	\includegraphics[width=0.8\textwidth]{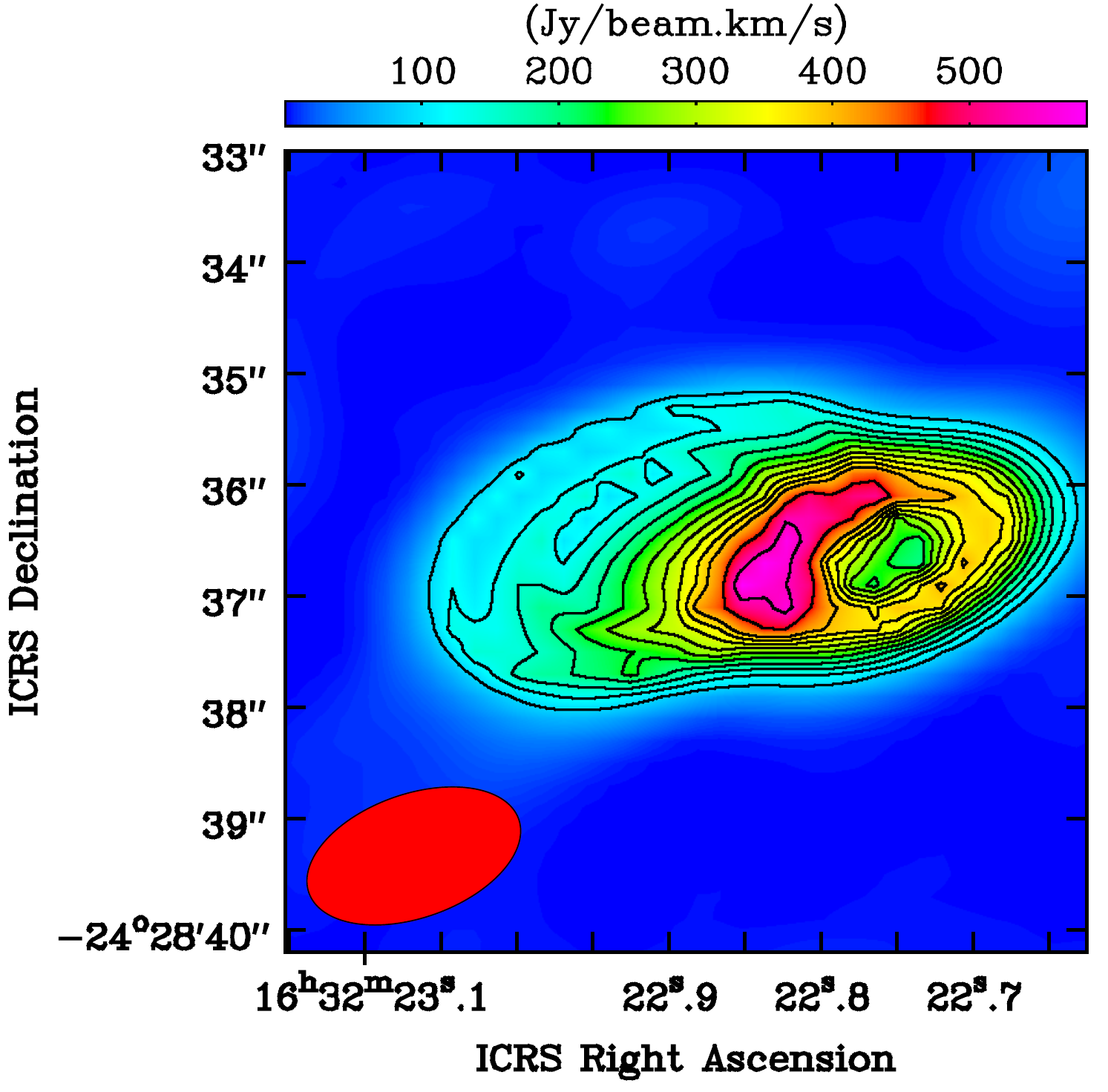}
	\caption{Integrated emission map of the observed \ce{H2O} at 183.310 GHz in the low mass protostar IRAS 16293--2422. The contour levels start at 10 mJy beam$^{-1}$ (3$\sigma$) and increase in the order of $\surd$2. The synthesized beam is shown in the lower left corner with red colour size of which is 1.99$^{\prime\prime}$$\times$1.11$^{\prime\prime}$.}
	\label{fig:emi}
\end{figure}
\section{Result and Discussion}
\label{res} 
\subsection{molecular emission line of \ce{H2O} in the low mass protostar IRAS 16293--2422 }
\begin{figure}
			\centering
	\includegraphics[width=0.8\textwidth]{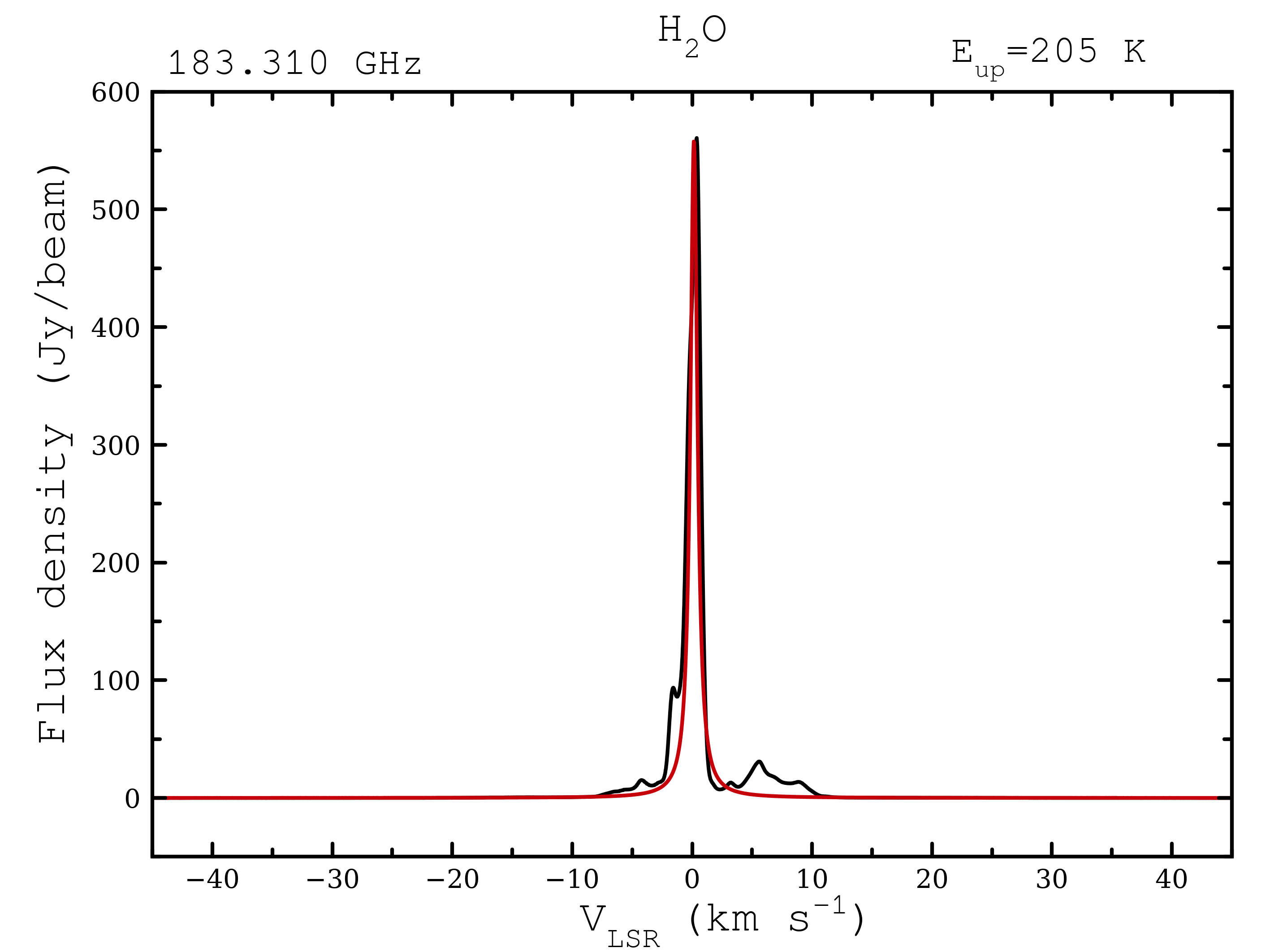}
	\caption{Disk average rotational emission spectrum of \ce{H2O} at frequency $\nu$ = 183.310 GHz in the low mass protostar IRAS 16293--2422. In emission spectrum, the black line shows the observed transitions of \ce{H2O}, while the red line shows the synthetic spectra obtained from the best LTE fits model which performed by the {\tt MADCUBA-AUTOFIT} tool to derive the proper column density of \ce{H2O} in the IRAS 16293--2422.}
	\label{fig:spec}
\end{figure}

We detect the rotational emission line of water vapour at the frequency $\nu$ = 183.310 GHz with transition J=3$_{1,3}$--2$_{2,2}$ and upper energy 205 K in the low mass protostar IRAS 16293--2422. The integrated emission map of \ce{H2O} towards IRAS 16293--2422 is shown in Fig.~\ref{fig:emi} and disk average rotational emission spectrum of \ce{H2O} is shown in Fig.~\ref{fig:spec}. The spectral peak of the water emission line was verified using the online {\tt Splatalogue}\footnote{\href{https://splatalogue.online//}{https://splatalogue.online//}} database for astronomical molecular spectroscopy. Recently, \citet{ko16} detected the emission line of \ce{H2O} at frequency 183.310 GHz and 325 GHz in Arp 220 using ALMA.

\begin{table*}
	\centering
	\caption{Properties of fiting parmeters in the emission line of water and value of column density are found after the fitting of LTE model.}
	\begin{adjustbox}{width=1\textwidth}
		\begin{tabular}{|c|c|c|c|c|c|c|c|c|c|c|}
			\hline 
			Species&Frequency&E$_{u}$&Beam Size&FWHM&Area&Column density&T$_{ex}$&Relative abundance \\
			& [GHz]  &[K] & [arcsec]  &[km s$^{-1}$]&[kJy beam$^{-1}$ km s$^{-1}$]&[cm$^{-2}$]&[K]&[\ce{H2O}/\ce{H2}]\\
			\hline
			\ce{H2O}&183.310&205&1.997$^{\prime\prime}\times$1.110$^{\prime\prime}$&1.284$\pm$0.006 &0.481$\pm$0.002 &4.2$\times$10$^{16}$&124$\pm$10&1.44$\times$10$^{-7}$\\
			\hline
		\end{tabular}
	\end{adjustbox}
	\label{tab:table}
\end{table*}

\subsection{Derivation of excitation temperature and column density of \ce{H2O}}
\label{sec:photo} 
We use the {\tt MADCUBA\footnote{\href{https://cab.inta-csic.es/madcuba/}{https://cab.inta-csic.es/madcuba/}}(ImageJ)} software toolkit to study and analyze the detected \ce{H2O} emission line which was built at the center for Astrobiology in Spain. This software is used to find the molecular transition of a detected species and derives the physical parameters such as molecular column density ($N_{tot}$), excitation temperature ($T_{ex}$), and linewidth ($\Delta$V) using spectroscopic data from CDMS and JPL databases. After that, the MADCUBA-Spectral Line Identification and modeling tool generate synthetic spectra in each detected line, based on current Local Thermodynamical Equilibrium (LTE) and line opacity effects. We manually adjust the FWHM = 1.284$\pm$0.006 km s$^{-1}$ to align the synthetic spectra to the observed line profiles. The {\tt MADCUBA-AUTOFIT} fitting tool is used to have the best nonlinear least squared fit using the Levenberg–Marquardt\footnote{\href{https://people.duke.edu/~hpgavin/ce281/lm.pdf}{https://people.duke.edu/~hpgavin/ce281/lm.pdf}} algorithm to reduce the $\chi^{2}$ value. The LTE fitting spectrum of \ce{H2O} is shown in Fig.~\ref{fig:spec}. In Tab.~\ref{tab:table}, we present the proper derived value of column density of water emission line with other fitting parameters.

\section{Summary}
\label{sec:discussion} 
In this letter, we present the spectroscopic detection of rotational emission line of \ce{H2O} in the low mass protostar IRAS 16293--2422 using ALMA band 5 observation. The emission line of water is found at the frequency $\nu$ = 183.310 GHz with $\geq$5 $\sigma$ statistical significance. The derived statistical column density of the emission line of \ce{H2O} using LTE model is $N$(\ce{H2O}) = 4.2$\times$10$^{16}$ cm$^{-2}$ with excitation temperature ($T_{ex}$) = 124$\pm$10 K. The relative abundance of \ce{H2O} with respect to \ce{H2} i.e., \ce{H2O}/\ce{H2} is 1.44$\times$10$^{-7}$ where column density of \ce{H2} is $N$(\ce{H2}) = 2.9$\times$10$^{23}$ cm$^{-2}$. 

In gas-phase chemistry, water (\ce{H2O}) is generated by the ion-molecule reactions that contribute to \ce{H3O}$^{+}$. It can dissociatively recombine to create \ce{H2O} \citep{bat86,rod02}. Water can also be created in protostars by the reaction O+\ce{H2}$\longrightarrow$OH + H, which follows by the reaction of OH with \ce{H2} \citep{wag87,hol89,atk04}. The water (\ce{H2O}) is formed in the dense and cold region on the grain surface of protostar by the sequence of reactions in which hydrogen and oxygen are accepted in the form of gas \citep{tie82,jon84,mok09,dul10}. The grain temperature rises above $\sim$100 K near protostars. As a result, the \ce{H2O} ice will go through desorption \citep{ce96,fra01} and an increase of \ce{H2O} gas-phase abundance in the inner part of the envelope of protostars \citep{mel00}. The detection of water vapour in the low mass protostar gives keys for the further studies of the formation mechanism of the water and deep studies of the water emission in the other star formation regions.

\section*{acknowledgements} 
	This paper makes use of the following ALMA data: ADS/JAO.ALMA\#2017.A.00042.T. ALMA is a partnership of ESO (representing its member states), NSF (USA), and NINS (Japan), together with NRC (Canada), MOST and ASIAA (Taiwan), and KASI (Republic of Korea), in co-operation with the Republic of Chile. The Joint ALMA Observatory is operated by ESO, AUI/NRAO, and NAOJ. The data that support the plots within this paper and other findings of this study are available from the corresponding author upon reasonable request. The raw ALMA data are publicly available at \href{https://almascience.nao.ac.jp/asax/}{https://almascience.nao.ac.jp/asax/}.

\end{document}